\documentclass{emulateapj}
\usepackage{graphicx}
\usepackage{amssymb}
\usepackage{amsmath}
\makeatletter

\usepackage{graphics,epsfig,lscape,longtable,rotating}

\slugcomment{}
\shorttitle{Chandra observation of J1834-087 }
\shortauthors{Misanovic, Kargaltsev \& Pavlov}

\makeatother

\begin{document}

\title{Chandra observation of the TeV source HESS J1834--087 }

\author{Zdenka Misanovic$^1$, Oleg Kargaltsev$^2$ and George G.\ Pavlov$^{3,4}$}

\affil{$^1$School of Physics, Monash University, Melbourne, 3800 VIC,
  Australia (Zdenka.Misanovic@monash.edu)\\
$^2$Dept. of Astronomy, University of Florida, Gainesville, FL
32611-2055 (oyk100@astro.ufl.edu) \\
 $^3$Dept.\ of Astronomy and Astrophysics,
The Pennsylvania State University, \\
525 Davey Lab., University Park,
PA 16802 (pavlov@astro.psu.edu)\\ 
$^4$St. Petersburg State Polytechnical University, Polytechnicheskaya ul. 29,
St.\ Petersburg 195251, Russia }

\begin{abstract}

 {\sl Chandra} ACIS observed the field of the extended TeV source 
HESS~J1834--087 for 47 ks.
A previous {\sl XMM-Newton} EPIC observation of the same field  revealed a 
  point-like source
(XMMU~J183435.3--084443) and an offset region of  faint extended emission.
  In the low-resolution, binned  EPIC images 
 the two appear to be connected.
However, the  high-resolution   {\sl Chandra} ACIS images  do not support the
 alleged connection.
 Instead, in these images XMMU~J183435.3--084443 is resolved into a point
 source,  
  CXOU~J183434.9--084443 ($L \simeq 2.5\times 10^{33}$ ergs
s$^{-1}$, for a distance of 4 kpc; photon index $\Gamma\simeq1.1$),
 and a compact
 ($\lesssim 20''$) nebula
with an isotropic morphology  and a
softer spectrum ($L\simeq4.2\times 10^{33}$ ergs s$^{-1}$, $\Gamma\simeq2.7$). The nature of the nebula is
uncertain. We discuss a dust scattering halo and a pulsar-wind nebula as
possible interpretations. Based on our analysis of the X-ray data, we
re-evaluate the previously suggested interpretations of  HESS~J1834--087 and
discuss a possible connection to the {\sl Fermi} LAT source  1FGL J1834.3--0842c. We also obtained an upper limit  of $3\times 10^{-14}$ ergs cm$^{-2}$ s$^{-1}$  on the unabsorbed flux of the SGR~J1833--0832 (in quiescence), which happened to be in the ACIS  field of view.

\end{abstract}


\section{Introduction}
\label{introduction}

Surveys  of the 
 Milky Way plane 
 with modern 
 Imaging Atmospheric Cherenkov Telescopes  (IACTs) 
have revealed more than 70 sources of very high energy
(VHE)
$\gamma$-rays\footnote{See http://tevcat.uchicago.edu/}.
While X-ray binaries (XRBs), young stellar clusters and background AGNs have
all been proposed   as possible sources of emission in the TeV energy range,
the majority of the  classified Galactic $\gamma$-ray sources are  thought to
be associated with  pulsar wind nebulae (PWNe) and supernova remnants (SNRs)
\citep{2005Sci...307.1938A,2006ApJ...636..777A,2008AIPC..983..195G,2010AIPC.1248...25K}.
However, about 30\% of the cataloged VHE sources lack reliable classifications.  Furthermore, for many classified 
 sources there remains an uncertainty about the actual mechanism responsible
 for the production of the 
  VHE  $\gamma$-rays. Therefore,
 multiwavelength studies and modeling of the VHE source population
 are of a great interest at the present time.

 The two main mechanisms for the production of high-energy $\gamma$-rays are 
the inverse Compton (IC) scattering of low-energy photons, and the hadronic
collisions, in which the energetic protons colliding with ambient particles
produce neutral pions, which decay into high-energy photons \citep[$\pi^0
\rightarrow 2\gamma $; see, e.g.,][]{2009ARA&A..47..523H}. Rotation-powered
pulsars emit highly energetic pulsar wind (PW), which is responsible for the
synchrotron radiation observed from radio up to MeV energies \citep[see][for 
recent reviews]{2008AIPC..983..171K,2010AIPC.1248...25K}. It is believed that
 the
$\gamma$-ray PWN can be produced when the ambient microwave, optical or infrared photons are upscattered by  relativistic PW electrons  up to the GeV and TeV energies.  
 While the X-ray PWNe are powered 
 by  energetic electrons   injected/accelerated in the vicinity of the pulsar (Lorentz factor $\sim
10^8$),  the  VHE PWNe are powered by cooled, less-energetic electrons (Lorentz
factor $\sim 10^7$), which have propagated further away from the pulsar \citep[e.g.,][]{2008arXiv0803.0116D}. 
    The IC cooling time of these
$\gamma$-ray-producing electrons is longer than that of the electrons
that produce  X-rays, and hence, they  trace the
cumulative history of the PWN, similar to radio emission (the relic PWN
model).   If the pulsar contains a hadronic component, the VHE emission could
also be produced  via a hadronic process when relativistic PW hadrons
encounter dense material such as the host SNR shell. An
alternative possibility is that the SNR shock accelerates 
 protons, which then interact with the dense ISM (e.g., a molecular cloud) and
 produce VHE emission.

The extended source HESS J1834--087 
\citep[$18.7\pm2.0\times 10^{-12}$ photons cm$^{-2}$ s$^{-1}$ 
 at $E>200$ GeV,
photon index $\Gamma = 2.5\pm 0.2$;][]{2005Sci...307.1938A,2006ApJ...636..777A} is spatially
 coincident with the center of the radio shell SNR G23.3--0.3 \citep[W41;][]{1992AJ....103..943K}, making the remnant a promising candidate for generating the observed $\gamma$-ray emission. However, the smaller  size of the HESS source\footnote{See Table 3 and Section 5.1 in \citet{2006ApJ...636..777A}.}  
 (diameter $\sim 10'$
 compared to the SNR diameter of $\sim 30'$) and the projected location well within the SNR shell suggest that the $\gamma$-ray emission does not come from 
 the entire SNR shell as it has been seen in other SNRs firmly associated
 with HESS sources \citep[e.g.,][]{2009A&A...505..157A}.
  \citet{2006ApJ...643L..53A} detected the TeV source with the {\sl Major
  Atmospheric Gamma Imaging Cerenkov (MAGIC) telescope},
 confirmed its
extension, and measured the flux 
 consistent with the HESS
measurement.

The region was also observed in
 radio with the Very Large Array (VLA), as part of the Galactic
Plane survey in continuum and in CO molecular line emission, and also in
X-rays with {\sl XMM-Newton}. Based on these
 observations,
\citet{2007ApJ...657L..25T} estimated the remnant's age ($10^4 - 10^5$ yrs)
and the distance (4 kpc), and  suggested that
 the observed $\gamma$-ray
emission is produced by a hadronic process when the particles
 accelerated by the
SNR shock interact with a giant molecular cloud
 (GMC) 
 projected near  the center of
W41  \citep[see also][]{2006MNRAS.371.1975Y}.  Another explanation was put
forward by \citet{2008MNRAS.385.1105B}, who
 proposed the nearby pulsar
J1833--0827 (=B1830--08; 24$'$ away from the center of the HESS 1834--078)
 as the source of
relativistic electrons. These electrons were
 ejected at the pulsar's birth
place, presumably near the SNR center, i.e., within the extent of the HESS
source (a relic PWN). \citet{1995MNRAS.275L..73G}
 argued that the association
 of the PSR J1833--0827 and W41 is plausible, given the pulsar's
 characteristic age of 148 kyr and the corresponding projected
 velocity of
 $\sim$250 km s$^{-1}$, required to move the
 pulsar from
its birthplace to the current location.  
However, in a recent analysis
 of the {\sl XMM-Newton} data,  Schmitt et al. (2010, to be submitted to ApJ) rules out the association of B1830--08 and W41. Schmitt et al. detected a possible PWN surrounding the pulsar, but since the morphology of the nebular emission does not resemble a bow-shock, they concluded that the pulsar does not move fast enough to travel from the center of W41 to its current position, approximately 26$'$ away.

Subsequent  analysis of the {\sl XMM-Newton}
 observations
 \citep[][hereafter M+09]{2009ApJ...691.1707M}
revealed a bright point-like source XMMU J183435.3--084443
 at the center of W41 apparently connected to
a region of faint extended emission also
 located within the extent of the HESS
source. 
M+09 have argued that the new pulsar/PWN candidate is the most likely counterpart of HESS~J1834--087.

To discriminate between these alternative scenarios, we have observed the region of the TeV source with the {\sl Chandra} ACIS detector. 
 In Section~\ref{observations} we present the multiwavelength
 data on HESS J1834--087, which,
  in addition to the {\sl Chandra} observation, also include archived radio,
  infrared (IR) and $\gamma$-ray data.  We further discuss our results in
  Section~\ref{discussion} and summarize them in Section~\ref{conclusion}.

\section{Observations and results}
\label{observations}

We observed the region around  HESS J1834--087 on 2009 June 7 with {\sl
  Chandra} ACIS (ObsID 10126) for 47.5 ks (46.5 useful scientific exposure)
 in timed
exposure (TE) mode using very faint (VF) telemetry format.  The point source
was imaged on the S3 chip approximately 15$''$ away from the nominal aimpoint,
placing a large part of the extended emission detected by {\sl XMM-Newton} on
the same chip. In addition, S1, S2, S4, I1 and I2 chips were also activated
during the observation. We used CIAO 4.2 and CALDB 4.2.2 for our analysis. We
have applied additional particle background cleaning for the VF mode, using
the pulse heights in the outer 16 pixels of the  $5\times5$ event island to help distinguish between ``good'' X-ray events and ``bad'' events that are most likely associated with cosmic rays. 
We also made extensive use of the archival multiwavelength data of this region obtained with {\sl XMM-Newton}, VLA and
 {\sl Spitzer} to compare them with the  {\sl Chandra} images.

\subsection{Spatial analysis}
\label{spatial}

\subsubsection{Previous multiwavelength observations}

\begin{figure}
\includegraphics[height=6.5cm,angle=0]{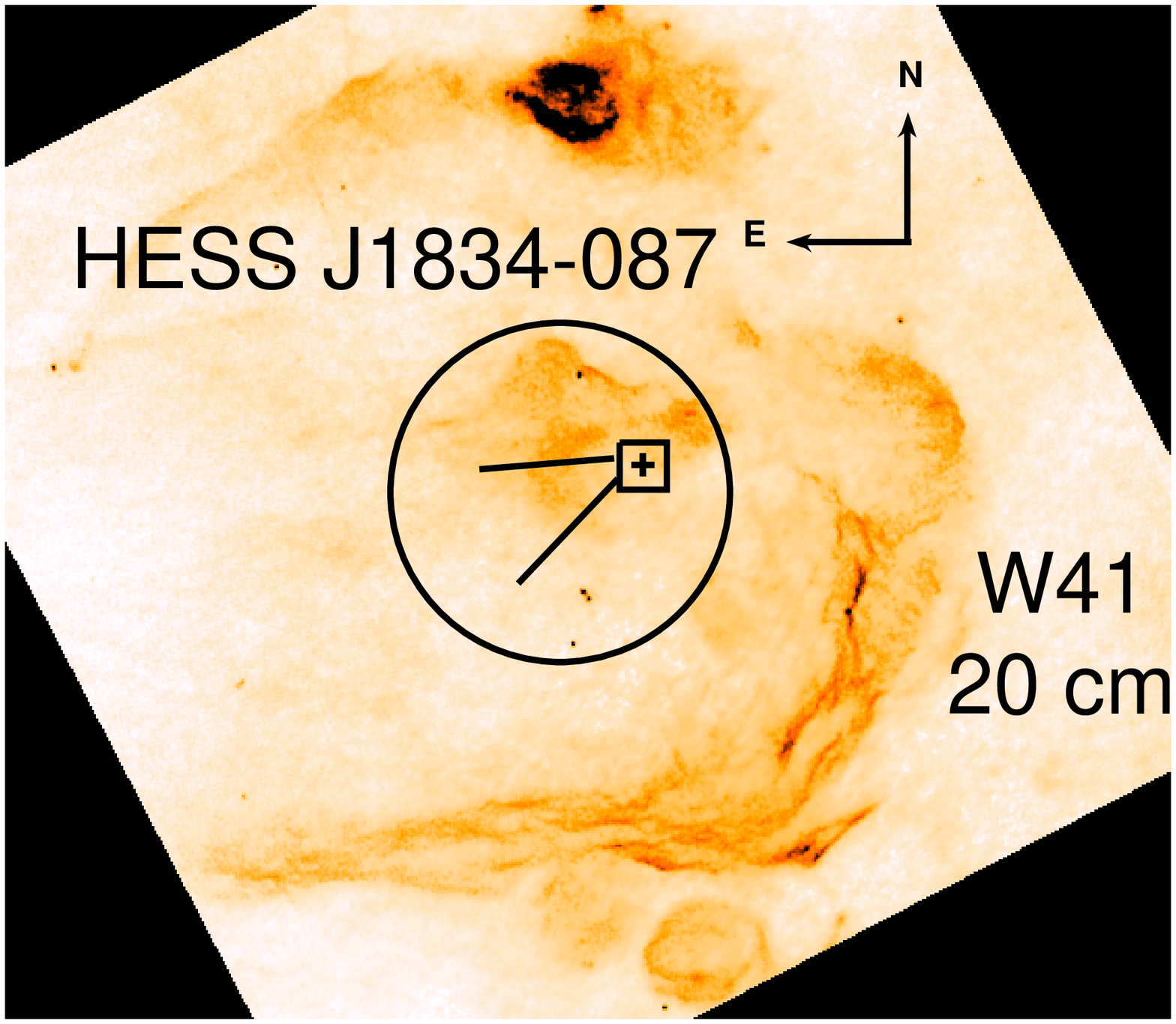}
\includegraphics[height=6.5cm,angle=0]{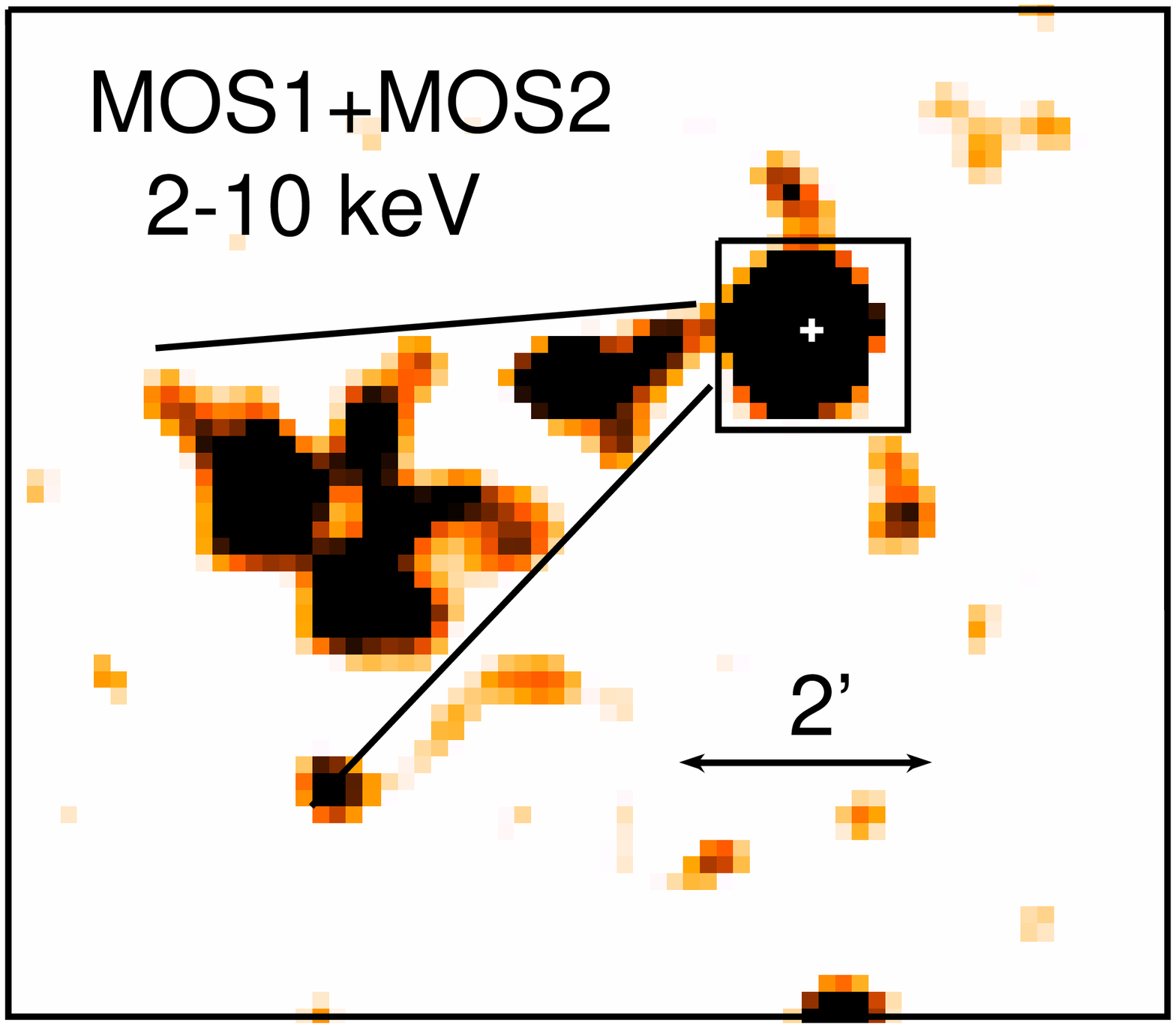}
\includegraphics[height=6.5cm,angle=0]{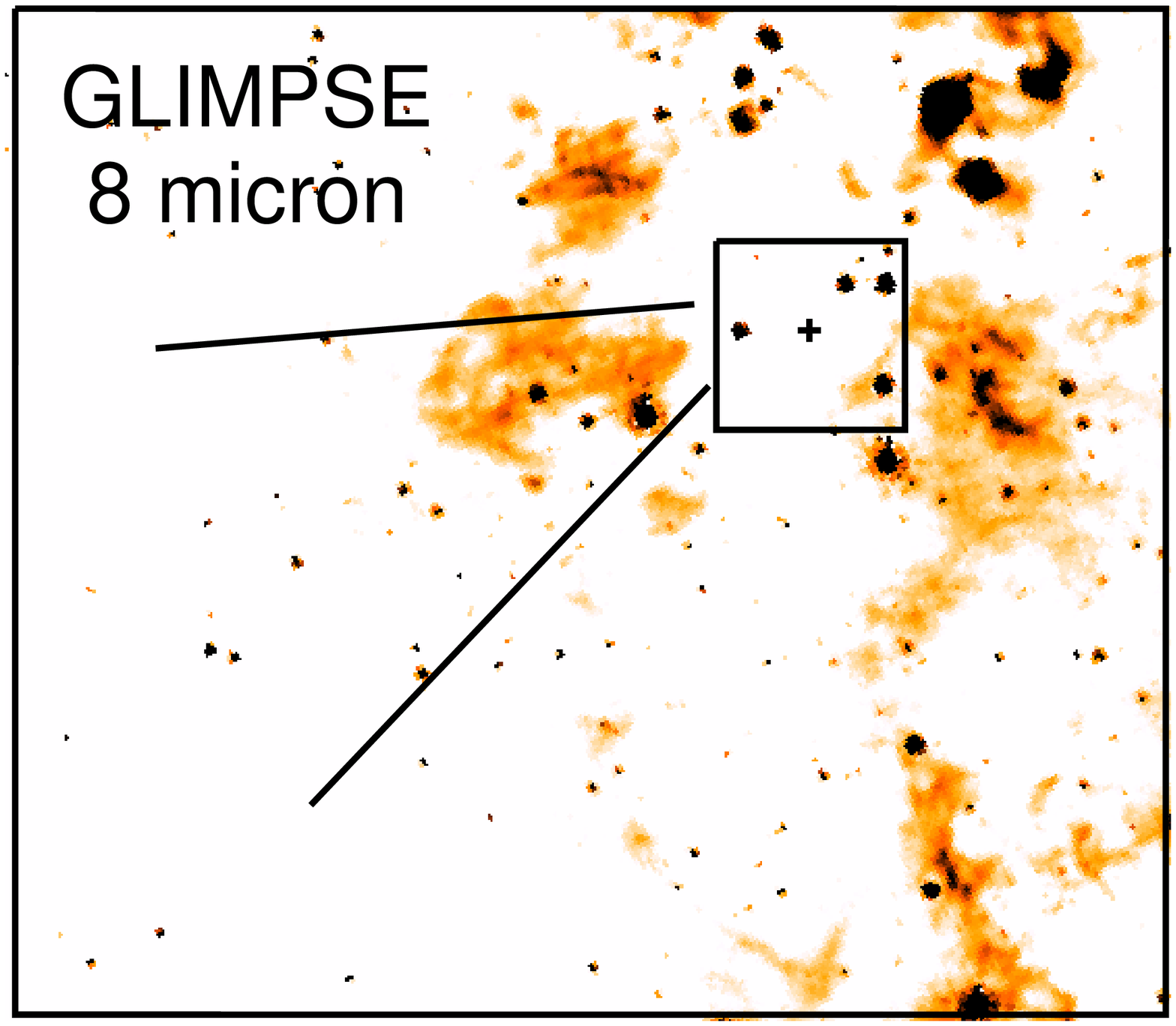}
\caption{{\sl Top:} 
VLA 20-cm image of the SNR G23.3--0.3 
(W41) showing its highly structured partial shell with a diameter of $\sim$30$'$ and some enhanced emission
near the center (White et al. 2005). The circle (radius 5\farcm4) shows the
position and extent  of HESS J1834--087 as measured by  \citet{2006ApJ...636..777A}.
{\sl Middle:} Broad-band (0.2--10 keV)  {\sl XMM-Newton} 17 ks  image
 of the interior of W41
 combining  MOS1 and MOS2 data (see also M+09). The image with the pixel size of 8$''$ is  smoothed with a Gaussian of FWHM 40$''$. 
The cross marks the position of CXOU~183434.9--084443 (see text).  Higher-resolution images of
 the vicinity of CXOU~183434.9--084443
 are shown
in Figure~\ref{fig-small}.  
The tail-like emission detected in the {\sl XMM-Newton} observation  is 
marked in all panels.  
 {\sl Bottom:} GLIMPSE (8 micron) image of the same region of the sky
 as in the middle panel.
\label{fig-large} }
\end{figure}

Figure~\ref{fig-large} shows the 
VLA 20-cm image of W41 (White et al. 2005),
 revealing fine details of the
  highly structured
 partial shell with a diameter of $\sim$30$'$, and also
 some extended radio emission
near the center.
 However, no point radio sources are visible, and no detection of radio pulsations from within the SNR shell has been  reported so far.
The 
extended VHE source HESS J1834--087 is 
 located near the SNR center. 
 With an angular 
extent of  5\farcm4 in radius \citep[see Table 3 in][]{2006ApJ...636..777A}
  the TeV source
appears 
 to be significantly
smaller than the SNR shell (Figure~\ref{fig-large}, left).

 The  middle panel of
Figure~\ref{fig-large} shows the innermost region of the extended TeV source
 as it is seen in a 
  17 ks  exposure by 
{\sl XMM-Newton} EPIC  \citep[for the analysis of {\sl XMM-Newton}
data see][M+09]{2007ApJ...657L..25T}.  
M+09 detected a total of 16 point-like sources in the FoV of the MOS
 detectors, with one of the brightest sources well within the extent of the 
 TeV emission. Furthermore, in the EPIC images this bright point-like source (XMMU~J183435.3--084443) appears to be
 accompanied by extended X-ray emission, also located within the extended TeV 
 source. 
 The spectral analysis of the {\sl XMM-Newton} data indicates that 
 both the point-like source and diffuse emission (designated XMMU
 J183435.3--084443 and G23.234--0.317, respectively; M+09) are highly 
 absorbed and most probably  non-thermal, although the spectral parameters for the  diffuse emission 
 are not well 
 constrained due to the large background contribution. 
  GLIMPSE\footnote{The Galactic Legacy Infrared Mid-Plane Survey; see http://www.astro.wisc.edu/sirtf} 8$\mu$m image (Figure~\ref{fig-large},
 bottom) shows no correlation between the diffuse X-ray and IR emission.

\subsubsection{Images} 
\label{images}

\begin{figure}
\includegraphics[height=6.5cm,angle=0]{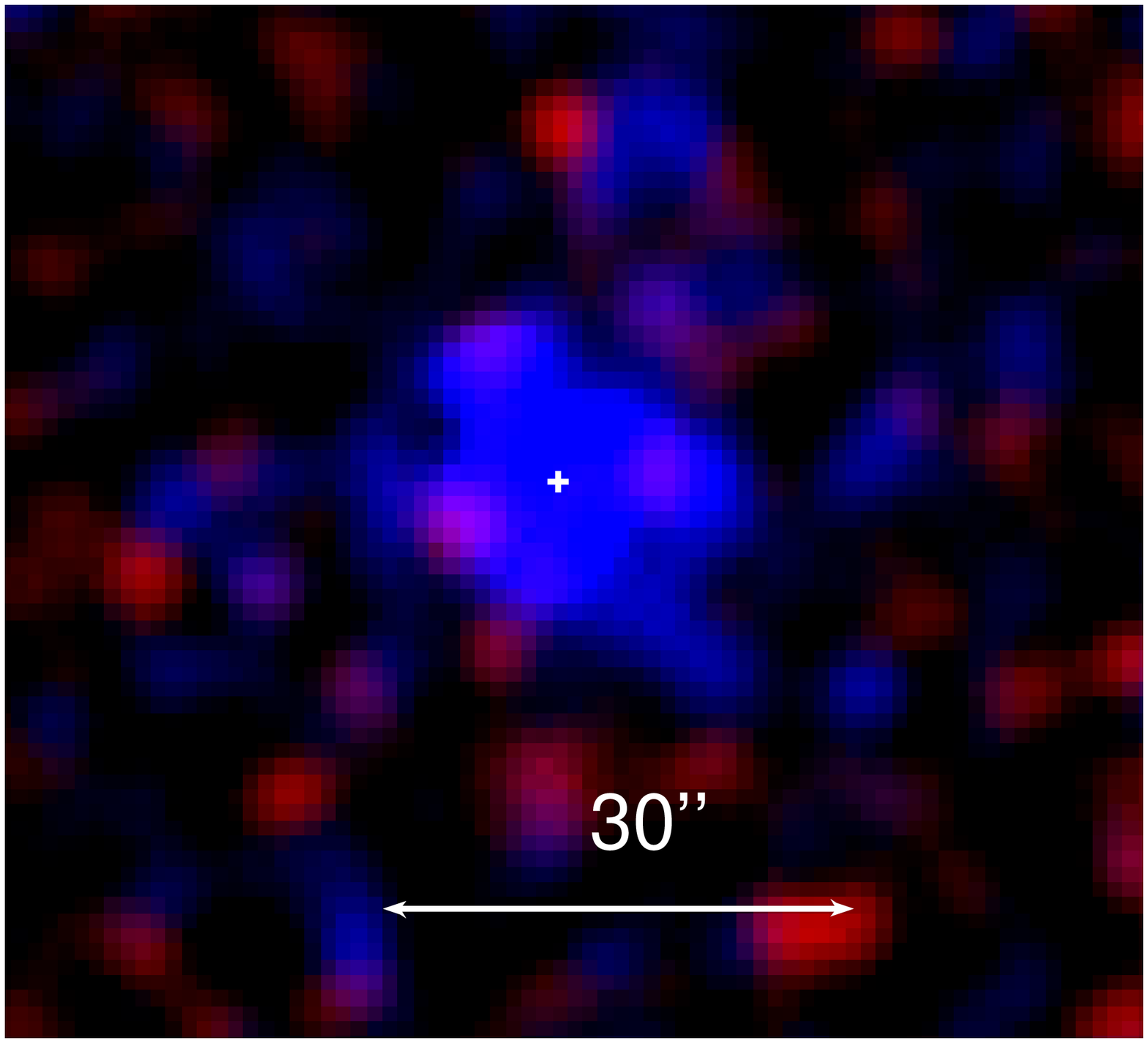}
\includegraphics[height=6.5cm,angle=0]{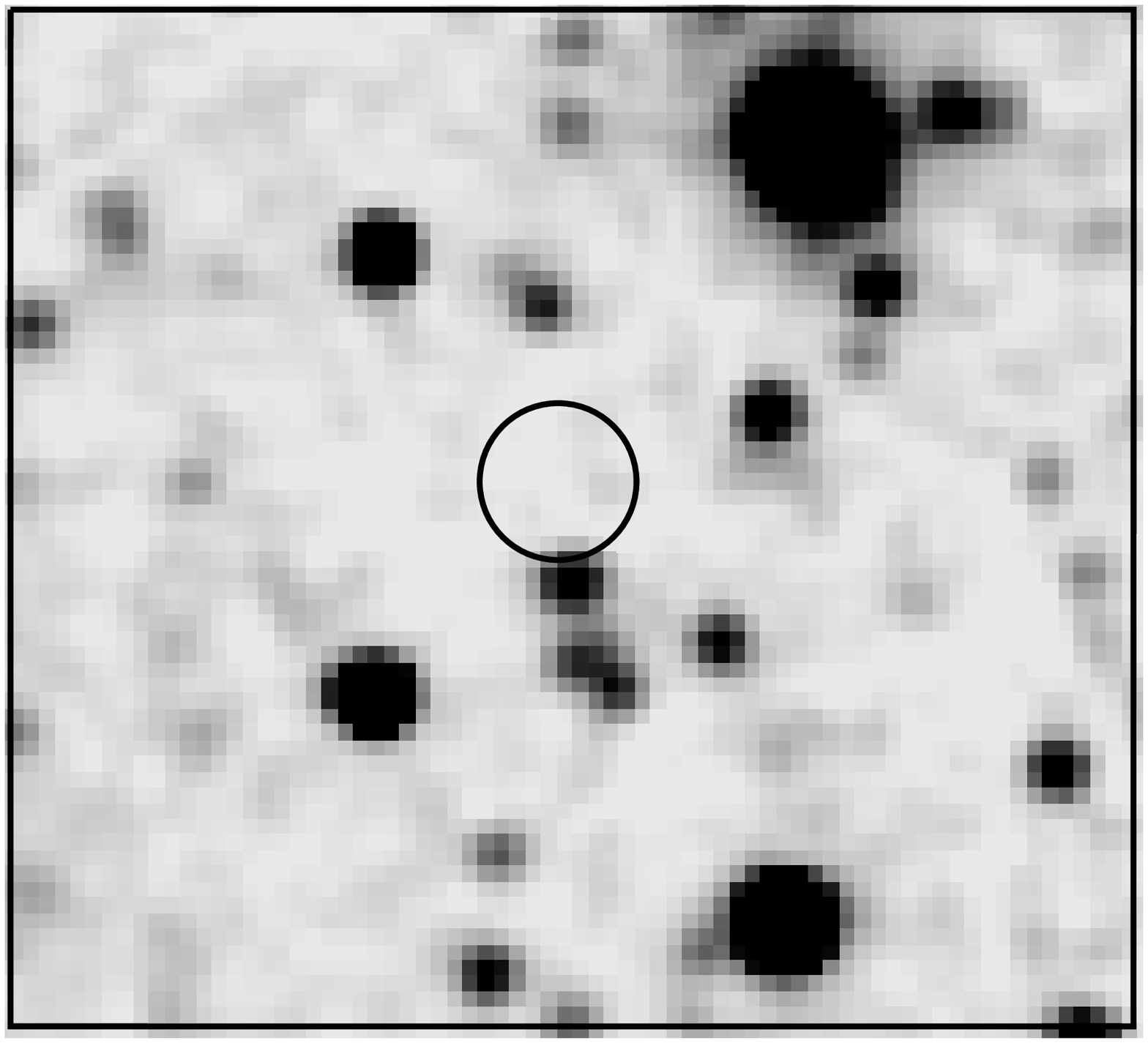}
\caption{{\sl Top:} {\sl Chandra} ACIS $90''\times 90''$ two-color (red: 0.5--2 keV; blue: 2--8 keV)  image of the region around
  the pulsar and compact PWN candidate coinciding with HESS
  1834--078. The  image is binned to a pixel size of 1$''$ and smoothed with a Gaussian of
  FWHM $\sim 4''$.  {\sl Bottom:}
 2MASS image of the same
  region. The position of the pulsar candidate is marked by a circle with a
  radius of 5$''$.  
 \label{fig-small} }
\end{figure}

Our {\sl Chandra} observation resolved XMMU~J183435.3--084443 into a point
 source, which we designate CXOU~183434.9--084443 (centered at R.A.=18$^{\rm
   h}$  34$^{\rm m}$ 34$^{\rm s}$.944, Decl.=--08$^o$ 44$'$ 43\farcs09), 
and a  
 compact diffuse emission extending up to $\simeq20''$ 
from the 
point source  (Figure~\ref{fig-small}, top),
 with  more or less isotropic  surface brightness distribution,
 and an average surface brightness of 
 0.8 cts arcsec$^{-2}$ in an annulus with radii 1\farcs5 and 10$''$
 (S/N$\approx$14). 
To separate the X-ray diffuse emission from the PSF wings  of the
pulsar candidate image,
 we have  simulated a  PSF, using MARX\footnote{For the MARX simulation we
   used the same off-axis and roll angles, and the detector parameters as in
   the real observation, but increased the exposure time by a factor of 100
   (rescaling the resulting image) to
 reduce the statistical errors. We found that using the dither blur parameter
 of 0\farcs35 gives the best match between the data and simulation.} and assuming the measured
 spectral properties of CXOU~J183434.9--084443.
  The comparison of the data and point
 source simulation (Figure~\ref{fig-profile}) suggests a good agreement between the data and simulation
 in a small aperture  (approximately up to 1$''$ radius) while the  
 extended emission  is easily seen at larger radii.
The 2MASS image of the same region (Figure~\ref{fig-small}, bottom) shows no
 near-infrared (NIR)
sources within $6''$ from CXOU~J183434.9--084443 position.


We searched for other point sources in the ACIS image in particular within the extent of the HESS source. We found a faint ($\sim$30 cts in 0.5--8 keV) point-like source  at 
 R.A.=18$^{\rm h}$  34$^{\rm m}$ 42$^{\rm s}$.6, Decl.=--08$^o$ 45$'$ 01\farcs95, approximately 2$'$ east of the pulsar candidate (Figure~\ref{fig-acis-tail}, top). The source coincides with the region of the enhanced ``tail'' emission detected by M+09, although no point source was detected at that position in the {\sl XMM-Newton} data.
  This could be due to  the lower angular resolution and higher EPIC background or due to the transient nature of the source. 
 The {\sl Chandra} spectrum of this source is soft, with an absorption column
 almost an order of magnitude lower than that of the candidate
 pulsar/PWN. Furthermore, the X-ray source coincides with a 2MASS source,
 suggesting that it is a foreground star. Most of the point-like X-ray sources
 detected in the {\sl XMM-Newton} data (Table 1
 of M+09)  are either outside the FoV of the activated ACIS
 chips, far off-axis, or near the chip edges, but we did detect the bright
 sources  11 and 15 on I2 and I3 chips, respectively (following the
 source designation by M+09).  The X-ray spectra of these sources are relatively soft, with absorption columns $N_{\rm H} = (1-9) \times 10^{21}$ cm$^{-2}$, suggesting that they are relatively nearby.  
 Source 15 coincides with a faint 2MASS source, already listed in Table 1 of  M+09. Using our more accurate {\sl Chandra} position, we also find two bright 2MASS sources within the {\sl Chandra} error circle of the source 11.
We also 
 found that the position of the recently detected
  SGR~J1833--0832 \citep{2010GCN.10526....1G,2010ApJ...718..331G} is 
  within the ACIS FoV, near the edge of the S1 chip (Figure~~\ref{fig-acis-tail}, bottom), but we did not detect any source at this position.
 After correcting for the vignetting, and assuming the spectral parameters reported by \citet{2010GCN.10531....1G}, we obtained an upper limit  of $3\times 10^{-14}$ ergs cm$^{-2}$ s$^{-1}$  on the unabsorbed flux of the SGR in quiesence.

 While looking for high-energy sources within the ACIS FoV, we 
 found  {\sl Fermi {\rm LAT}} source 1FGL J1834.3--0842c
\citep[0.11 Crab in 1--100 GeV, photon index
$\Gamma = 2.24\pm 0.04$;][]{2010ApJS..188..405A} 
 located $\sim$5$'$ northwest of
CXOU~183434.9--084443. The 95\% error ellipse of the 1FGL J1834.3--0842c
 position 
also includes  the northwestern part of the SNR
shell (Figure~~\ref{fig-acis-tail}, bottom).
The only X-ray source  
detected within  
 the error ellipse of the {\sl Fermi {\rm LAT}} source is a
relatively faint source, CXOU~183430.3--084142.8, with a bright optical counterpart, located at the edge
of the GeV emission.  This source is apparently a  foreground
star, most likely not associated with the GeV emission.

Figure~\ref{fig-acis-tail} (top) shows the large-scale ACIS image, which covers the region of the extended HESS source.
The heavily smoothed image  shows some faint extended emission,
 which must be a  
part of the large-scale emission seen in
 the {\sl XMM-Newton} images. Indeed, a closer comparison of the two images 
(see Figure~\ref{fig-large}) shows that
the extended emission visible in the ACIS image coincides with the brightest regions seen in the {\sl XMM-Newton} data.
 However, the origin of this emission remains elusive.
 We see no clear connection between CXOU~183434.9--084443 and
the extended  emission $3'-4'$ east-southeast of it. Its association with the SNR is also
 questionable, as it does not
 seem to coincide
  spatially with the radio shell or with the diffuse radio emission in the SNR
  interior  (Figure~\ref{fig-acis-tail}, middle and bottom). To investigate the nature of this extended emission, its spectrum must be measured in a long
 observation.

 XMMU~J183435.3--084443 has also been detected by the {\sl Swift} \citep{2006ApJ...651..190L}. M+09 included the three archived 
 {\sl Swift} observations in their analysis, but could not make any firm conclusions about intrinsic flux variability of XMMU~J183435.3--084443  (the count rate measured  in the third observation showed a 3$\sigma$ increase over the first two, or 2$\sigma$ increase over the mean rate).
To search for variability in the ACIS observations, we produced light-curves
with a bin size ranging from $\sim 10^2$ to $10^3$ seconds, but could not see
any significant flux variations on the time-scales relevant for the ACIS data
(i.e., considering the length and time resolution of the ACIS observation).
Our {\sl Chandra} data allowed us to search for pulsations in the 6--100
s period range, but we did not find any periodic signal.

\subsection{Spectral analysis}

\begin{figure}
\includegraphics[height=8.5cm,angle=0]{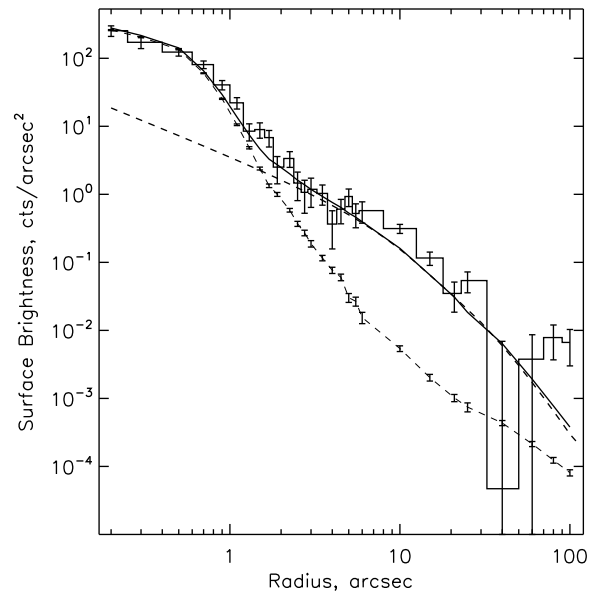}
\caption{
   The radial profiles (in 3--8 keV) of the simulation and data showing the contribution of
   the extended emission in the vicinity of the point source. The dashed line with errorbars 
   shows the PSF profile of the point source simulated using MARX (see
   text). The histogram shows the radial distribution of the surface brightness measured in
   annular  regions  centered on the point source. The background level of 0.15 cts arcsec$^{-2}$ has been
    subtracted from the
     data.   The dashed line without errorbars shows possible contribution of
     a dust halo (see Section~\ref{dust-halo} and
 Appendix, equations A6 and A7). 
 \label{fig-profile} }
\end{figure}

\begin{figure}
\includegraphics[height=6.5cm,angle=0]{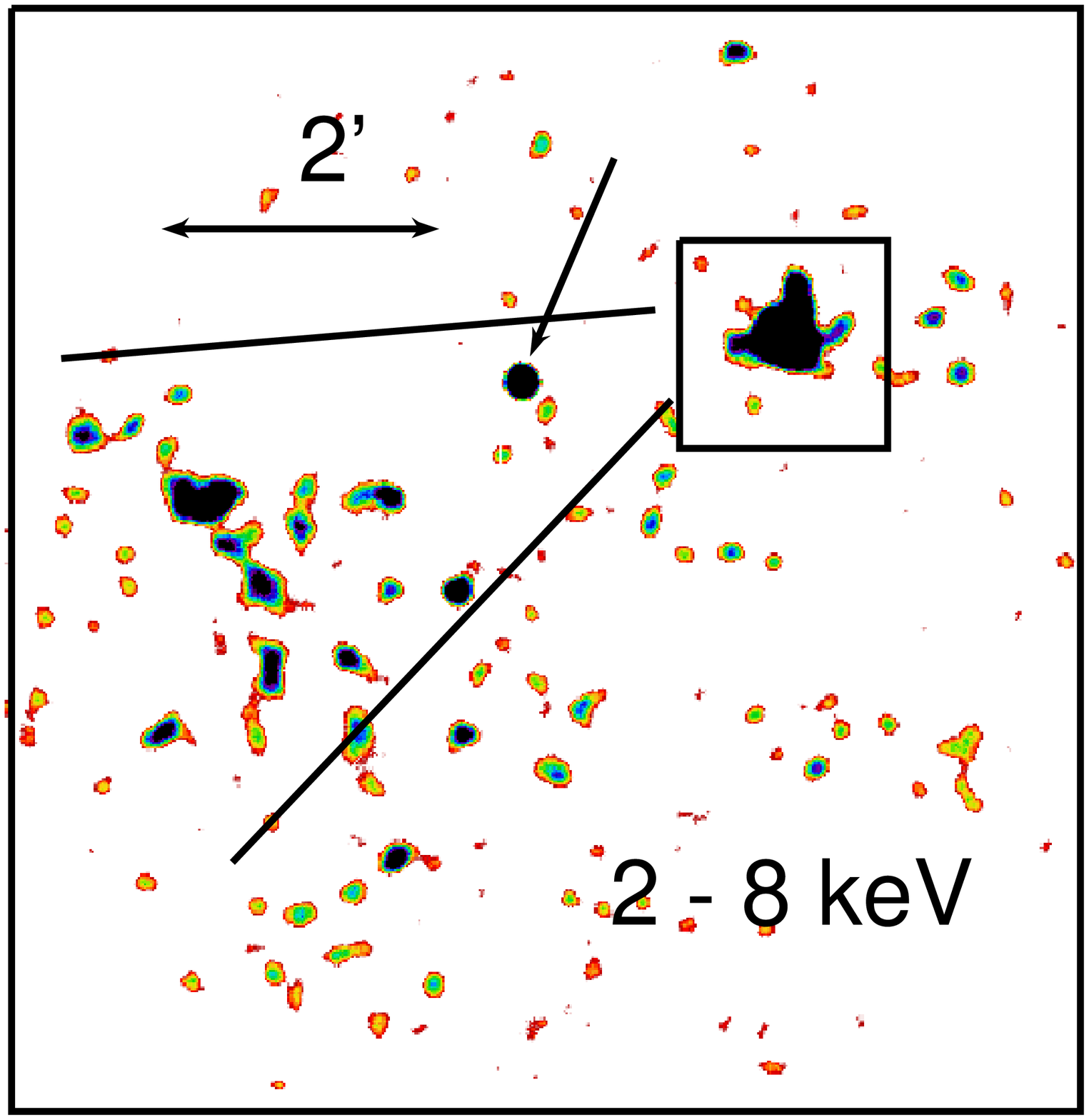}
\includegraphics[height=6.5cm,angle=0]{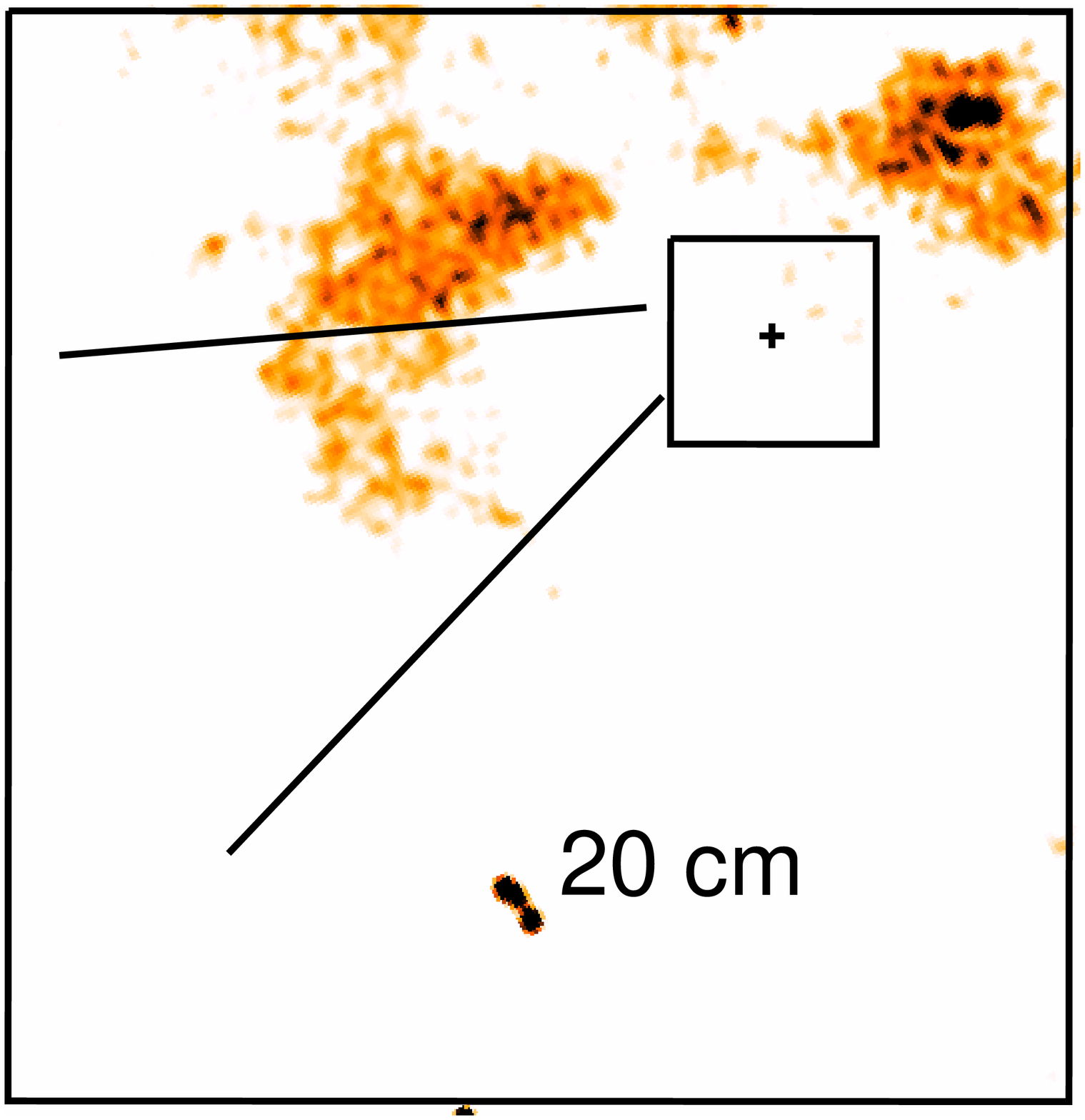}
\includegraphics[height=6.5cm,angle=0]{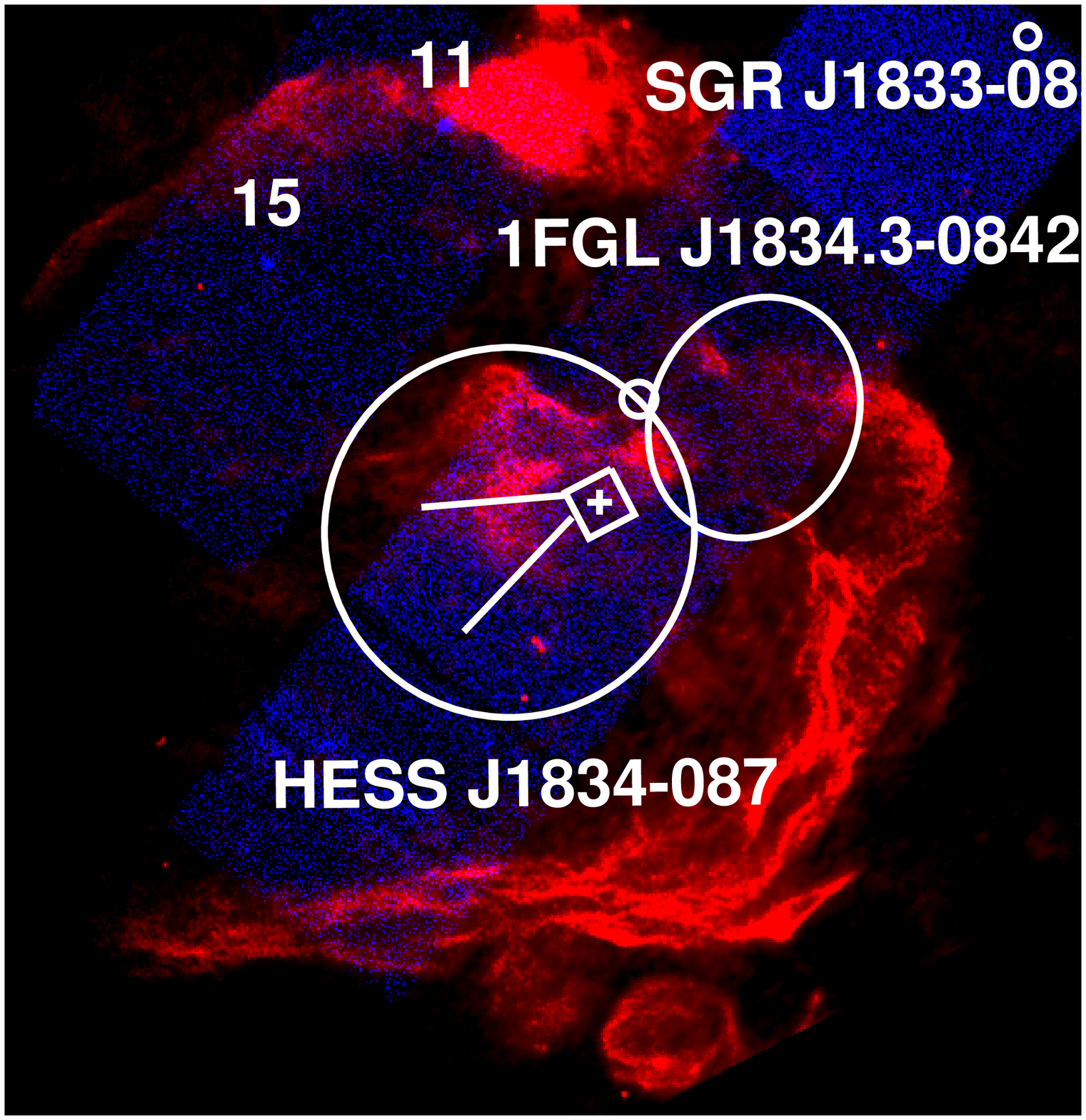}
\caption{{\sl Top:} Large-scale, smoothed with a Gaussian of FWHM$\sim10''$, {\sl Chandra} ACIS image (2--8 keV)
  encompassing  the region of the extended  
  tail-like emission detected
  with {\sl XMM-Newton} (see Figure~\ref{fig-large}). The arrow shows the
  point source with a bright 2MASS counterpart (see text). {\sl Middle:} VLA
  image of the same region.   The enhanced extended radio emission in the
  center of W41 does not seem to coincide with the diffuse X-ray
  emission. {\sl Bottom:}
A two-color image (red: 20 cm; blue: 0.5-8 keV)
  showing the whole W41 and also the projected ACIS chips activated in the
  {\sl Chandra} observation. The extent of the TeV source is marked by the big
  circle (radius 5\farcm4)  while the GeV source is shown by
  its 95\% positional error ellipse (6\farcm0$\times$7\farcm2). Several X-ray sources detected in
  the FoV are marked, including the  star candidate at the edge of the GeV
  emission (small circle), and sources 11 and 15 as numbered in Table~1 by M+09. The position of the SGR J1833--0832
 is also marked with a small circle.   
 \label{fig-acis-tail} }
\end{figure}

To minimize the contribution from the
 nebula, we extracted the X-ray spectrum of 
  CXOU~J183434.9--084443 from a circular
region with a radius of 1$''$  centered on the point source (see Figure~\ref{fig-profile}), while the
background photons were extracted from an annulus with the inner and outer
radii of 1\farcs5 and 5\farcs5, respectively.
 A total of 220 counts were detected
in the  $r = 1''$ aperture in the 0.5--8 keV band. The comparison with the
simulated point source suggests that approximately 30 of these counts are from
the nebula, 
 while the contribution from the background is negligible
(1--2 counts), implying an aperture-corrected\footnote{From the radial profile
  of the simulated PSF and the data (Figure~\ref{fig-profile}) we estimated
  the point source contribution of  85\% in this aperture.} count rate of
$4.80\pm0.04$ cts ks$^{-1}$.  For spectral fitting these counts were grouped to a minimum of 15 counts per bin.

To measure the nebular spectrum, we extracted a total of 208 photons
from the 1\farcs5--10$''$ annulus  centered on CXOU~J183434.9--084443, while the background contribution was measured from an annulus with $10''< r < 20''$.
 We estimate that there are $\sim$50 counts from the true background and $\sim$10 counts from the PSF wings in the same aperture,
implying a count rate of $3.2\pm1.1$ cts ks$^{-1}$ for the background- and
PSF-subtracted emission.
 The nebula spectrum was also grouped to a minimum of 15 counts per bin.

\begin{figure}
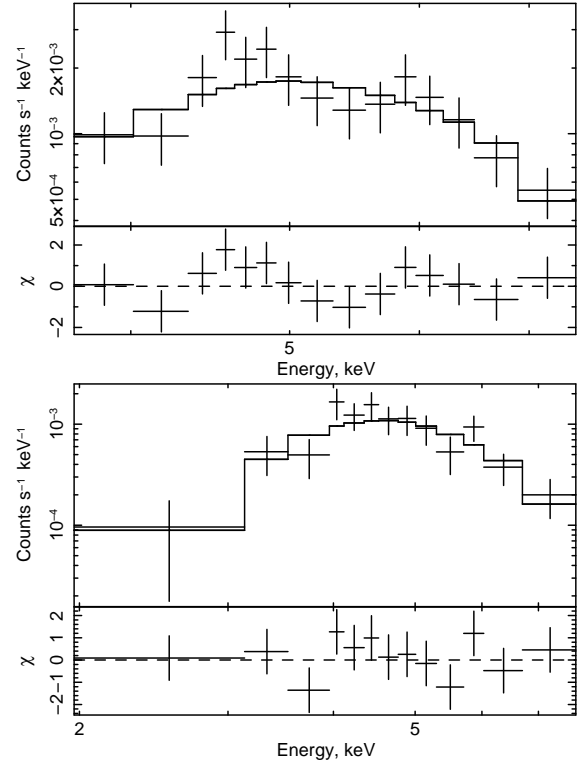

\includegraphics[height=7.5cm,angle=-90]{fig5a.ps}
\includegraphics[height=7.5cm,angle=-90]{fig5b.ps}
\caption{{\sl Top:} The X-ray spectrum of the pulsar candidate extracted from the {\sl Chandra} observation,
  and the corresponding best-fit absorbed PL model. The
  absorption column is frozen to its best fit-value (see
  Table~\ref{table-spectrum}) determined from the fit combining the
 {\sl Chandra}
  and {\sl XMM-Newton} data (see text). {\sl Bottom:} The X-ray
  spectrum of the PWN candidate extracted from the {\sl Chandra} data and the
  correponding absorbed PL model, assuming the same absorption column as for
  the pulsar candidate (Table~\ref{table-spectrum}).    
 \label{fig-spectrum} }
\end{figure}

\begin{figure}
\includegraphics[height=8.5cm,angle=90]{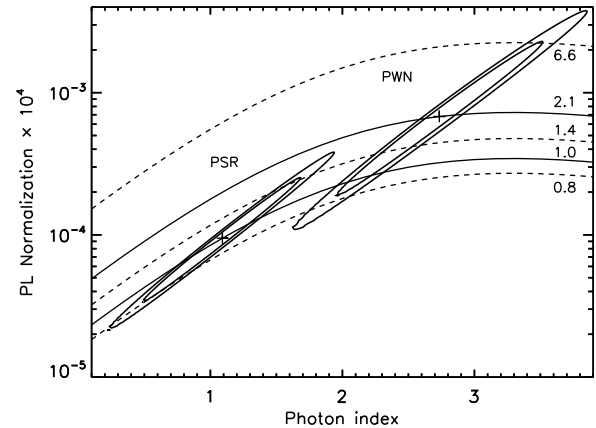}
\caption{90\% and 99\% confidence contours and the unabsorbed fluxes
  (0.5--8 keV) in units of $10^{-12}$ ergs cm$^{-2}$ s$^{-1}$ for the pulsar
  and PWN candidate.   
 \label{fig-contours} }
\end{figure}

The small number of counts and the lack of soft emission  (the spectra are 
virtually cut off below 
3 keV; see Figure~\ref{fig-spectrum})
 did not allow us
to constrain well the spectral model parameters. 
  We fit the spectra with absorbed power-law (PL) models and show the
  results in Table~\ref{table-spectrum} and Figures~\ref{fig-spectrum} and \ref{fig-contours}. We first fit each spectrum
  independently and noticed that the absorption columns for the point source
  and for the diffuse emission region were similar ($2.8^{+2.8}_{-1.3}$ and
  $2.8^{+2.9}_{-1.5}$ $\times 10^{23}$ cm$^{-2}$, respectively), suggesting
  a common distance. 
 To determine the value of the absorption column more accurately, we then
 performed a simultaneous fit with the MOS1 and MOS2 spectra\footnote{The MOS
   spectra were only used to measure the absorption column, because they contain not only photons from the pulsar candidate but also a large fraction of the compact nebula that could not be resolved.} of 
 XMMU~J183435.3--084443, assuming again a common absorption column density
 (i.e., 
linking this parameter when fitting the spectra simultaneously, with other parameters untied).

After obtaining the best-fit value for $N_{\rm H}$, we fixed this parameter
 before fitting the ACIS spectra of  the point source and the nebula 
  individually and calculating the errors for
 other model parameters and for the fluxes (Table~\ref{table-spectrum}).   
  The unabsorbed luminosities are calculated for the 0.5--8\,keV band assuming a distance of 4 kpc \citep{2007ApJ...657L..25T}.
We note that the large absorption measured from the X-ray spectra is more than
an order of magnitude higher than the estimated $N_{\rm HI}$ value for
the Galaxy in that direction \citep[$\sim2\times 10^{22}$
cm$^{-2}$;][]{1990ARA&A..28..215D}, implying a large amount of molecular
material in the line of sight \citep[see][and references
therein]{2007ApJ...657L..25T}.

\begin{table*}
\scriptsize
\caption{
Absorbed PL model fits to the spectra of CXOU~J183434.9--084443 and  surrounding  extended emission.}
\begin{center}
\begin{tabular}{cccccccc}
\hline\hline\\
  & $N_{\rm H}$ & $\Gamma$ & PL Norm. & $\chi_{\nu}^2$/d.o.f & Absorbed Flux  & Luminosity  \\
 &$10^{23}$ cm$^{-2}$ & & $10^{-4}$\,cm$^{-2}$\,s$^{-1}$\,keV$^{-1}$ & & $10^{-13}$\,ergs\,cm$^{-2}$\,s$^{-1}$ & $10^{33}$\,ergs\,s$^{-1}$ \\
\hline \\

Point source & 2.72 (frozen) &  $1.09^{+0.40}_{-0.42}$ &  $0.95^{+0.92}_{-0.47}$ &
0.81/13 &$2.56^{+0.22}_{-0.22}$ & $2.3^{+0.6}_{-0.4}$  \\

 & & & & & & & \\

Nebula & 2.72 (frozen) & $2.73^{+0.54}_{-0.53}$ &  $6.8^{+9.1}_{-3.9}$ &
0.64/13 & $0.98^{+0.13}_{-0.12}$ & $4.1^{+4.8}_{-1.9}$ \\
& & & & & & & \\
\hline\\

\end{tabular}
\end{center}
\tablecomments{
 The absorption column density was frozen at its best-fit value determined in
 the joint fit with {\sl XMM-Newton} data (MOS1 and MOS2 spectra of the
 point-like source XMMU~J183435.3--084443).
The observed flux and unabsorbed luminosity, 
$L_X=4\pi d^2 F_X^{\rm unabs}$,
 are calculated for the 3--8\,keV and 0.5--8\,keV energy band, respectively,
 for the distance $d=4$ kpc.
 The listed uncertainties are at a 1$\sigma$ confidence level. The observed flux and unabsorbed luminosity of CXOU~J183434.9--084443 are aperture-corrected. 
}
\label{table-spectrum}
\end{table*}

\newpage

\section{Discussion} 
\label{discussion}

 {\sl Chandra} 
  resolved 
  the compact extended emission surrounding  CXOU~J183434.9--084443, which is by far the brightest X-ray source 
located within the extent  of 
 HESS J1834--087.
  It was previously suggested that the TeV emission 
 could arise 
 either as a result of interaction between the SNR shock and  the  GMC near W41 \citep{2007ApJ...657L..25T},  or alternatively,  it could  be due to a PWN powered by CXOU~J183434.9--084443  (M+09).  Below we discuss 
 the constraints that our observation places on the nature of CXOU~J183434.9--084443 and HESS J1834--087. We also explore a possible link between these sources  and the recently discovered  {\sl Fermi} LAT source 1FGL J1834.3--0842c which is also located within SNR W41.

\subsection{CXOU~J183434.9--084443 and its extended emission}

 The point source appears to have a strongly absorbed spectrum that fits hard PL. The flux, PL slope, and the lack or any counterparts at lower 
  frequencies make    CXOU~J183434.9--084443  a plausible candidate for a young remote pulsar. However, other options, such as an AGN seen through the Galactic plane, or a quiescent, obscured XRB (similar to those 
   discovered with {\sl INTEGRAL}), remain viable. Although the latter interpretation could explain the observed large $N_{\rm H}$ by the intrinsic absorption, 
  we do not find any evidence of variability or an iron line which are
  commonly  seen  in obscured HMXBs   \citep{2006A&A...453..133W}. It seems
  rather plausible that a large amount of   molecular   material is generally
  present along this line of sight; for instance,  \citet{2010ApJ...718..331G}
  found $N_{\rm H}\simeq 10^{23}$ cm$^{-2}$ while fitting the spectrum of the nearby SGR~1833--0832  in the active state.

An important piece of evidence, which we will consider in detail,  is a compact, $r\lesssim 20''$, extended emission,
 which has been clearly resolved from the point source (Figures~\ref{fig-small} and \ref{fig-profile}).  This emission appears to be approximately symmetric  
 and exhibits a much softer spectrum compared to that of the point source
 ($\Gamma=2.7\pm0.5$ vs.\ $1.1\pm0.4$). Two most likely  interpretations  are
 a PWN around a young pulsar in W41, or a dust scattering halo around the strongly absorbed point source.
  A combination of the two is also a possibility. In fact, a hint of a plateau  (or non-monotonic behavior) seen in Figure~\ref{fig-profile}  
  between $3''$--$10''$ may indicate at  two possible components with different dependence on $r$.

\subsubsection{A dust halo around  a point source?}
\label{dust-halo}

 The very large absorbing column measured in X-rays suggests a significant dust column that should  
  lead to  an extended dust scattering   halo
  \citep[e.g.,][]{1995A&A...293..889P}.   The halo brightness grows with the increasing intervening dust column, which is usually assumed to be proportional to the hydrogen absorption column, $N_H$, measured from  X-ray spectra.  
 More specifically,  the
scattering optical depth 
 $\tau_{\rm scat}\simeq S (N_H/10^{22}~{\rm cm}^{-2})(E/1~\rm{keV})^{-2}$ for $S\simeq0.5$ found by \citet{1995A&A...293..889P} 
 gives $\tau_{\rm scat} < 1$ at $E > 3.7$ keV  for CXOU~J183434.9--084443, whose
 spectrum virtually cuts off  below 3 keV
 (see Figure~\ref{fig-spectrum}). 
  Although  
 taking into account multiple scatterings would be more accurate, the large
 uncertainties of the other parameters (such as the dust grain properties and
 dust distribution; see Appendix)  do not  warrant this extra complication.
  Therefore, we
 have used the single scattering approximation to calculate the halo profile by convolving the spectral intensity of a halo  
   with the detector response in the 3--8 keV energy range.  We find that for
   the parameters $\Theta=360''$ and $S\simeq1$, and for the dust distribution
   function $f(x)$  defined in the Appendix (equation A8),
 the dust halo model
   generally describes the observed  radial profile (see
   Figure~\ref{fig-profile}). However, some deviations are noticeable, hinting
   at a possibility of a second emission
 component.
   The dust distribution function $f(x)$, used in the calculation of the
   halo model shown in Figure~\ref{fig-profile}, suggests that at least some
   dust must be located near the source, in agreement with the extreme absorption, which could be attributed to the local molecular clouds that are known to exist in this region of sky \citep{2006ApJ...643L..53A}.

We should, however, point out that,  in addition to the freedom of choosing the functional dependence  of dust distribution with the distance, the approximate  dust scattering model we use has two other  free parameters ($S$ and $\Theta$), which can attain 
 values within rather broad ranges, depending on the unknown  properties of the dust grains along
  this particular line of sight (see Appendix).  
  Therefore, the mere fact that the model qualitatively 
  describes the observed radial   profile does not guarantee that the extended emission is indeed a dust halo. On the other hand, the symmetric shape and the softer spectrum of the extended emission  support at least partial contribution of a dust halo.  Although the dust halo interpretation of the extended emission does not rule out the young pulsar option for the point source, it allows for additional possibilities,  such as a magnetar, an ANG, or an XRB.

\subsubsection{A  pulsar-wind nebula?}

 Let us now 
   assume that  CXOU~J183434.9--084443 is a pulsar, and a substantial fraction of the observed extended emission around it is a PWN.  The symmetric morphology of the candidate compact PWN would then suggest that the putative pulsar does not move very fast.  Even if the pulsar was born at the geometrical center of the SNR (i.e., $\sim 2'$ from its current position) $100$ kyrs ago, its transverse speed would only be  $22 d_4$ km s$^{-1}$, which
 is in agreement with the  nearly isotropic PWN shape.  Thus, the large-scale emission east of the pulsar 
 is not expected to be akin to long collimated tails
 formed behind supersonically moving pulsars, such as observed by
 \citet{2008ApJ...684..542K}. Indeed, the high-resolution X-ray images of
 those tails show that the surface brightness is usually the highest near the
 pulsar, and it gradually decreases with the distance from the pulsar, i.e., the opposite of what we see in this case.
 It could be, however, that the large-scale X-ray emission is akin to that in
 the Vela X PWN, where the X-ray emission is offset from the pulsar and thought to come form the relic pulsar wind 
  crushed and pushed aside  by the reverse SNR shock \citep[e.g.,][]{2008ApJ...689L.121L}.

   The slope of  of the extended emission  spectrum, fitted with the absorbed PL model, is
  rather steep, $\Gamma=2.7\pm0.8$, albeit uncertain. For a PWN, such  a steep slope is   unusual because the X-ray spectra of most\footnote{A recently discovered PWN
    candidate in HESS J1632--408  \citep{2010A&A...520A.111B}  also shows a steep spectrum with
    $\Gamma=3.4^{+0.6}_{-0.8}$. However, the authors  did not consider a possible contribution of a dust halo, which may be quite substantial given the very large $N_{\rm H}\simeq1.3 \times 10^{23}$ cm$^{-2}$ they measured in X-rays. } PWNe  have  $\Gamma=1-2$ \citep[e.g., see Figure 6 in][]{2008AIPC..983..171K}. The steep
  slope may, however, be indicative of  strong synchrotron cooling.    The
   ratio of the extended and point source luminosities,  
 $L_{\rm PWN}/L_{\rm PSR}\sim 1.8$, is typical  of PWNe
 \citep[see Figure 5
in][]{2008AIPC..983..171K}.     The spin-down power of the putative pulsar powering the PWN can be estimated as $\dot{E}=4\pi r_{s}^2c p_{\rm amb}$, where $r_{s}$ is the termination shock radius and $ p_{\rm amb}$ is the ambient pressure. In the Sedov expansion phase, the pressure inside the SNR could be estimated from 
 its radius by using a simple formula $p_{\rm amb}=3E/4\pi R^3$, where it is
 usually assumed that the total SNR explosion energy $E=10^{51}$ ergs
 \citep[e.g., see][]{2009A&A...496....1D}. The estimated pressure $p_{\rm
   amb,-9}= p_{\rm amb}/10^{-9}~ {\rm dyne~cm}^{-2}$  and the termination
 shock radius scaled to a plausible value  $r_{s,17}= r_{s}/10^{17}~{\rm cm}$
 (corresponding to 2.7$''$ at the W41 distance of $d_4=4$~kpc) give
 $\dot{E}=4\times10^{36} r_{s,17}^2 p_{\rm amb,-9}$ erg s$^{-1}$ -- a value
 typical for a young Vela-like pulsar. This estimate is sensitive to the value
 of  $r_{s,17}$, which could be a factor of a few smaller then the scaling
 chosen\footnote{Tori radii in bright, well-resolved PWNe exceed $r_s$ by a
   factor of 2 on average \citep{2010ApJ...709..507B}. }.  The luminosity of
 the compact PWN, $\simeq4.1\times10^{33}d_{4}^2$ erg s$^{-1}$, could also be
 used to estimate $\dot{E}$ from the $L_{X}$--$\dot{E}$ correlation
 \citep[see][for recent results]{2008AIPC..983..171K}. Although this
 correlation shows a very large scatter, the plausible range for $\dot{E}\sim
 10^{36}-10^{37}$ erg s$^{-1}$ is in a good agreement with the above estimated value.

\subsection{The nature of HESS J1834--087, 1FGL J1834.3--0842c, and  CXOU~J183434.9--084443 }
  
   HESS J1834--087  belongs to the growing  group  of  Galactic TeV sources that lack firm classifications or identifications with   lower-energy counterparts.  There are currently about 25 TeV sources in this group, of which 
   only five can be considered truly ``dark'' sources (i.e.,\  those without
   any detected counterparts) while the rest of  the sources are  coincident with one or more  lower-energy 
    sources of a known nature (SNRs, massive open stellar clusters,  GMCs, X-ray binaries), which could, in principle, power the TeV emission (although there are no clear-cut cases).
  Spatial coincidence with W41 puts HESS J1834--087  in  the latter category. 
  As such a good positional match between  a TeV source and a bright radio SNR
  is very unlikely to happen by chance,  HESS J1834--087 must have physical
  relation to W41.  This naturally  leaves
  only  two options for the origin of HESS J1834--087, a relic PWN or an SNR
  shock interacting with the GMC, both of which have been previously discussed
  \citep[M+09,][]{2007ApJ...657L..25T}. 
 
 Our detection of extended emission around  CXOU~J183434.9--084443 does not
 provide a convincing argument in  favor of either of these two possible
 interpretations.  Proving that CXOU~J183434.9--084443 is a pulsar (e.g., by
 detecting pulsations) would provide a very strong support to the relic PWN hypothesis.  Despite our  highly  significant detection of extended emission around CXOU~J183434.9--084443,  it 
   is difficult to unequivocally establish its origin 
 because of the very high absorption and a high likelihood  of a significant  dust scattering halo.  It appears that a better way to go about establishing the nature of  CXOU~J183434.9--084443 would be 
 a sensitive timing search 
 in X-rays and GeV. 
    However, even if CXOU~J183434.9--084443  turns out to be a pulsar, there will remain a question about the origin of the nearby   {\sl Fermi {\rm LAT}} source 1FGL J1834.3--0842c  (see Figure~\ref{fig-acis-tail}), which is located within W41 but offset 
 from both the center of the TeV source and from CXOU~J183434.9--084443, and
 even more offset from the patch of the large-scale X-ray emission, which
 could be a relic PWN akin to the Vela X PWN 
  (see above). Unless the 1FGL source is not real\footnote{It is marked as
    ``confused'' in the 1FGL catalog, which means that it might be the result of an imperfect model of the diffuse Galactic background \citep[see][for details]{2010ApJS..188..405A}.} or its position is  inaccurate, a separate interpretation would be required.  While awaiting for better {\sl Fermi} image,   we can speculate that  1FGL J1834.3--0842c might be able to provide a long-sought evidence of  hadrons in a pulsar wind \citep[e.g.,][]{2009ASSL..357..373A,2007Ap&SS.309..179B}.  Indeed,  the positional coincidence with the SNR suggests that the GeV emission can  be produced by pulsar-wind 
protons interacting with  the dense material of the part of the SNR shell,
which intriguingly turns out to be the nearest to the putative pulsar. We must
admit, however,  that until the pulsar nature of  CXOU~J183434.9--084443 is
firmly established the above reasoning will remain highly speculative.

 On the other hand, a failure to find pulsations from  CXOU~J183434.9--084443,
 down to a restrictive limit for  a young pulsar, would strongly suggest that
 it is an unrelated source. In this case, one would be forced to conclude that
 there is no plausible candidate for a young pulsar within W41. This would
 imply that the TeV emission is most likely produced by the interaction of
 protons accelerated in the SNR shock with a particularly dense material in
 the nearby molecular cloud
 \citep[][]{2007ApJ...657L..25T,2006ApJ...643L..53A}. Obtaining a firm
 evidence of this  would be of a great interest because so far  there are very
 few TeV sources (HESS J1745--303A, HESS J1800--240AB, HESS J1848--018) where
 the interaction with the nearby molecular cloud is considered as a possible
 (albeit not firmly established) mechanism of the VHE emission. Since HESS
 J1834--087 is located well within the W41 shell and has a significantly
 smaller diameter, it could only be associated  with 
 a part of the shell projected within the W41 interior on the sky.  The offset between the TeV and 1FGL sources would be even more difficult to explain than in the PWN interpretation.  Also, one would have to conclude that the large-scale X-ray emission (see Section~\ref{images}) must be associated with the SNR interior, which would require rather high temperatures to explain the measured hard spectrum (M+09).

\begin{figure}
\includegraphics[height=9cm,angle=90]{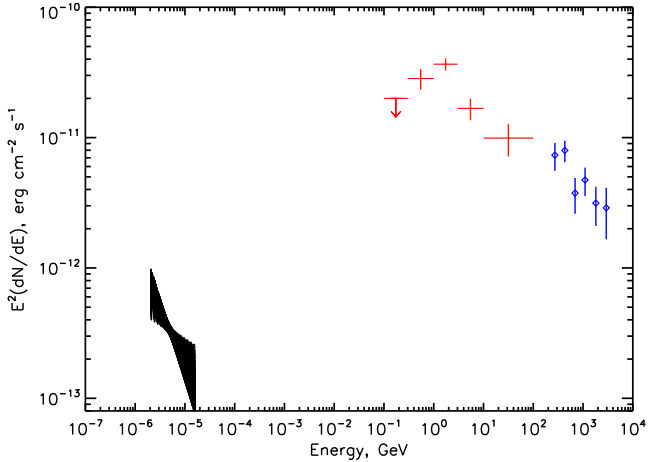}
\caption{Broad-band spectrum of the PWN candidate showing the X-ray spectrum
  (ACIS, black) and the catalogued GeV ({\sl Fermi}, red) and TeV (HESS, blue)
  emission. The TeV nebula spatially overlaps with the X-ray PWN candidate, while the GeV emission is located between the TeV nebula and part of the SNR shell (see Figure~\ref{fig-acis-tail}).
 \label{fig-sed} }
\end{figure}

Keeping in mind the  uncertainty in the interpretations,   we opt to plot the
multiwavelength PWN spectral energy distribution (Figure~\ref{fig-sed})
 but refrain from any modeling, which would be premature due to the large number of options afforded by the existing data.
 We note, however, that  the GeV and TeV
spectra generally seem to match, which hints at  their common origin and a possible inaccuracy in the determination of the 1FGL J1834.3--0842c position.
 Alternatively, the observed  GeV
emission coinciding with the  part of the shell could  be produced by
protons interacting with  hadrons in the SNR shell and  producing neutral pions
that decay emitting the observed $\gamma$-rays. The relativistic protons may
be accelerated  either in the PW
\citep[e.g.,][]{2009ASSL..357..373A,2007Ap&SS.309..179B}  or by the SNR
shock interacting with the molecular clouds, which has been also previously proposed as the source of the TeV emission \citep[][]{2007ApJ...657L..25T,2006ApJ...643L..53A}.

\section{Conclusion and summary}
\label{conclusion}

 Our {\sl Chandra } observation resolved the extended X-ray emission in the
 immediate vicinity of CXOU~J183434.9--084443, by far the brightest X-ray
 source within the HESS J1834--087 extent.  The source and the accompanying
 extended emission are strongly absorbed in X-rays leading to significant
 uncertainties in their spectra, which fit  PL models 
 with quite different slopes. The spectrum of the extended emission is
 markedly softer suggesting that the emission might be a dust scattering
 halo. Alternatively, the difference in the slopes could be due to the strong synchrotron cooling, if the extended emission is  a PWN.  Although our approximate  dust scattering  halo calculations allow us to obtain a reasonable  fit to the observed 
 radial profile of the extended emission, large uncertainties in the dust
 scattering model preclude us from definitive judgment on the nature of the
 extended emission, leaving a PWN option as a still viable alternative. The
 previously reported faint, large-scale extended emission appears to be
 disjoined from CXOU~J183434.9--084443 in a better resolution {\sl Chandra}
 image, which rules out a pulsar tail hypothesis and suggest that it might
 either be a part of a displaced relic PWN or a surprisingly hot plasma in the SNR interior.

Adjacent to the TeV source, and overlapping with a part of the SNR shell,
there is extended GeV emission, the {\sl Fermi {\rm LAT}} source 1FGL J1834.3-0842c.
 The offset from HESS J1834--087 and CXOU~J183434.9--084443 complicates the interpretation of the GeV emission suggesting that either 
 the 1FGL J1834.3-0842c's position is inaccurate or  it could be an intriguing
 case of a pulsar wind interacting with the SNR shell via hadronic mechanism.  
  The X-ray timing and  deeper imaging observations are required to understand
  the nature of  CXOU~J183434.9--084443 and of the offset large-scale X-ray
  emission.  Together with improved GeV data, this will make it possible to
  identify 
HESS J1834--087 and 1FGL J1834.3-0842c, and determine the nature of their relationship with W41.

We also serendipitously observed the recently discovered SGR J1833--0832,
which, however, was not detected.
We place a limit of  of $3\times10^{-14}$ ergs s$^{-1}$ cm$^{-2}$ on its unabsorbed flux, which is a factor of 40 dimmer than was measured by \citet{2010ApJ...718..331G}.

\acknowledgments
{Support for this work was provided by the National Aeronautics and Space
Administration through {\sl Chandra} Award Number GO9-0076X issued
by the {\sl Chandra} X-ray Observatory Center, which is operated by the
Smithsonian Astrophysical Observatory for and on behalf of the National
Aeronautics Space Administration under contract NAS8-03060. The work
was also partially supported by NASA grants NNX09AC84G and NNX09AC81G,
and NSF grants No.\ 0908733 and 0908611. G.~G.~P was partly supported by the
Ministry of Education and Science of the Russian Federation (contract 11.634.31.0001).}


\appendix 
\section{Dust halo model}

Dust halos, often seen around bright point-like X-ray objects, are formed by scattering
of source X-ray photons on dust grains. Here we will only discuss the case of
dust optically thin with respect to the
photon scattering, $\tau_{\rm scat}\lesssim 1$, and consider only
azimuthally symmetric halos
(which implies that the dust distribution across the line of sight (LOS) is uniform within the interval of angles $\theta$ at which we see the halo). In this case
the spectral halo intensity (ph\,cm$^{-2}$ s$^{-1}$ keV$^{-1}$ arcmin$^{-2}$) is given by the equation
\begin{equation}
I_{\rm halo}(\theta, E) = F(E) N_H \int_0^1 dx\, \frac{f(x)}{x^2}\, \frac{d\sigma_s(E,\theta_s)}{d\Omega_s}\,,
\end{equation}
where $F(E)$ is the point source spectral flux (photons\,cm$^{-2}$ s$^{-1}$ keV$^{-1}$), $x=(D-d)/D$ is the dimensionless distance from the X-ray source to the
scatterer ($D$ and $d$ are the distances from the observer to the source and the scatterer, respectively), $\theta_s\simeq \theta/x$ (for small angles) is the scattering angle, $f(x)$ is the dimensionless dust density distribution along
the LOS ($\int_0^1 f(x)\,dx =1$),
and $d\sigma_s(E,\theta_s)/d\Omega_s$ is the differential scattering cross section per one hydrogen atom, averaged over the dust grain distribution over sizes
and other grain properties (see, e.g., \citealt{1991ApJ...376..490M}).

To understand the halo properties from simple analytical expressions, we will
use the Rayleigh-Gans (RG) approximation, in which the total scattering
cross section $\propto E^{-2}$; this approximation works better for higher
energies, $E \gtrsim 0.5$--2 keV, depending on the dust model.
For some dust models, the averaged differential cross section in the RG
 can be approximated as (Draine 2003)
\begin{equation}
\frac{d\sigma_s(E,\theta_s)}{d\Omega_s} \approx \frac{\sigma_s(E)}{\pi \theta_{s,50}^2}\,
\frac{1}{(1 + \theta_s^2/\theta_{s,50}^2)^2}\, ,
\end{equation}
where
\begin{equation}
\theta_{s,50} \approx \frac{\Theta}{E}\quad {\rm and} \quad
\sigma_s(E) \approx \frac{S}{E^2}\,10^{-22}\,{\rm cm}^2
\end{equation}
are the median scattering angle and the total cross section, respectively;
$E$ is the energy in keV.
The constant $\Theta$ in first eq.\ (A3) depends on the dust model; \citet{2003ApJ...598.1026D}
derived $\Theta = 360''$ from the dust model of \citet{2001ApJ...548..296W}, while \citet{2010A&A...520A..71B} found $\Theta = 7.4'$ for the model of \citet{1998ApJ...503..831S}.

It follows from the second eq. (A3) that the scattering optical depth is
\begin{equation}
\tau_{\rm scat}(E) \approx S N_{H,22} E^{-2}.
\end{equation}
The factor $S$ in the second eq.\ (A3) is a constant of the order of 1;
e.g., $S\approx 1.3$ from Figure 6 of \citet{2003ApJ...598.1026D}, while \citet{1995A&A...293..889P} found a mean value $S\approx 0.49$ for a number of halos observed with {\sl ROSAT} (but the scatter was very large), while \citet{1991ApJ...376..490M}
discuss models with $S= 0.903$, 1.09, and 0.47 (see their Table 1).
Costantini (2004; PhD thesis) estimated $\tau_{\rm sca}(1\, {\rm keV})$ for a number of halo sources observed with {\sl Chandra}; the values of $S$ derived from her results show a very strong scatter, $S$ from 0.018 to 2.26. The scatter itself may be natural, as the dust properties may be different for different sources.

It should be noted that the correlation of $\tau_{\rm sca}(1\,{\rm keV})$
with visual extinction $A_V$:
\begin{equation}
\tau_{\rm scat}(1\,\,{\rm keV}) =
(0.056 \pm 0.01) A_V
\end{equation}
\citep{1995A&A...293..889P} is better than that with $N_H$, but $A_V$ is rarely known for the objects
of interest.

Substituting (A2) and (A3) in (A1), we obtain the spectral intensity profile
\begin{equation}
I_{\rm halo}(\theta, E) = F(E) N_{H,22} \frac{S}{\pi \Theta^2} \int_0^1 \frac{f(x)}{x^2}\,
\left[1 + \left(\frac{\theta E}{x \Theta}\right)^2\right]^{-2} dx\,.
\end{equation}

For comparison with the point source + halo profile observed in the energy
 range
$E_1 < E < E_2$, the sum of the spectral intensities should be convolved with
the detector response, with allowance for the image spread caused by the
telescope and the detector. We have checked that the energy redistribution in the detector only slightly affects the broadband radial profile for a smooth incident spectrum. Therefore, assuming the observable halo size to be much larger than the PSF width, we obtain
\begin{equation}
I_{\rm obs}(\theta) = \int_{E_1}^{E_2} dE 
A_{\rm eff}(E)\, F(E)\,\left\{\psi(\theta,E) +
N_{H,22} \frac{S}{\pi \Theta^2}\int_0^1 dx \frac{f(x)}{x^2}\left[1 + \left(\frac{\theta E}{x\Theta}\right)^2\right]^{-2}\right\} ,
\end{equation}
where $A_{\rm eff}(E)$ is the detector's effective area, and $\psi(\theta,E)$ is the normalized PSF, which can be taken from a simulation (e.g., with MARX). The first term in Equation (A7) corresponds to the point source, while the second term describes the halo.

The above equation can be integrated for a given set of halo parameters, and compared directly with the data. In particular, for the dust halo model shown in Figure 3 we picked 
 $\Theta=360\arcsec$, $S=1$, and the dust distribution  function
 \begin{equation}
f(x)=
\begin{cases}
\textrm{$x_o^{-1}$}, &\textrm{$x \leq x_o$}\\
0, &\textrm{$x > x_o$}
\end{cases}
\end{equation}
with $x_o = 0.25$. According to (A6), this distribution function corresponds to 
\begin{eqnarray}
I_{\rm halo}(\theta, E) = \frac{F(E) N_{H,22} S a}{2\pi (x_o \Theta)^2}
\left(\arctan a-\frac{a}{1+a^2}\right)
\nonumber \\  \approx  \frac{F(E) N_{H,22} S a}{4 x_o^2 \Theta^2}
\begin{cases}
1 &\textrm{$a \gg 1$} \\
\textrm{$(3\pi)^{-1} a^3$} & \textrm{$a \ll  1$}
\end{cases}
\end{eqnarray}
where $a=x_o \Theta / \theta E$.

\bibliographystyle{apj}
\bibliography{./paper}

\begin{thebibliography}{}

\bibitem[\protect\citeauthoryear{{Abdo} et~al.}{{Abdo}
  et~al.}{2010}]{2010ApJS..188..405A}
{Abdo}, A.~A., et~al. 2010, \apjs, 188, 405

\bibitem[\protect\citeauthoryear{{Acero} et~al.}{{Acero}
  et~al.}{2009}]{2009A&A...505..157A}
{Acero}, F., {Ballet}, J., {Decourchelle}, A., {Lemoine-Goumard}, M., {Ortega},
  M., {Giacani}, E., {Dubner}, G.,  \& {Cassam-Chena{\"i}}, G. 2009, \aap, 505,
  157

\bibitem[\protect\citeauthoryear{{Aharonian} et~al.}{{Aharonian}
  et~al.}{2005}]{2005Sci...307.1938A}
{Aharonian}, F., et~al. 2005, Science, 307, 1938

\bibitem[\protect\citeauthoryear{{Aharonian} et~al.}{{Aharonian}
  et~al.}{2006}]{2006ApJ...636..777A}
{Aharonian}, F., et~al. 2006, \apj, 636, 777

\bibitem[\protect\citeauthoryear{{Albert} et~al.}{{Albert}
  et~al.}{2006}]{2006ApJ...643L..53A}
{Albert}, J., et~al. 2006, \apjl, 643, L53

\bibitem[\protect\citeauthoryear{{Arons}}{{Arons}}{2009}]{2009ASSL..357..373A}
{Arons}, J. 2009, in Astrophysics and Space Science Library, Vol. 357,
  Astrophysics and Space Science Library, ed. {W.~Becker}, 373

\bibitem[\protect\citeauthoryear{{Balbo} et~al.}{{Balbo}
  et~al.}{2010}]{2010A&A...520A.111B}
{Balbo}, M., {Saouter}, P., {Walter}, R., {Pavan}, L., {Tramacere}, A., {Pohl},
  M.,  \& {Zurita-Heras}, J. 2010, \aap, 520, A111

\bibitem[\protect\citeauthoryear{{Bamba}, {Mori}, \& {Shibata}}{{Bamba}
  et~al.}{2010}]{2010ApJ...709..507B}
{Bamba}, A., {Mori}, K.,  \& {Shibata}, S. 2010, \apj, 709, 507

\bibitem[\protect\citeauthoryear{{Bartko} \& {Bednarek}}{{Bartko} \&
  {Bednarek}}{2008}]{2008MNRAS.385.1105B}
{Bartko}, H.,  \& {Bednarek}, W. 2008, \mnras, 385, 1105

\bibitem[\protect\citeauthoryear{{Bednarek}}{{Bednarek}}{2007}]{2007Ap&SS.309.%
.179B}
{Bednarek}, W. 2007, \apss, 309, 179

\bibitem[\protect\citeauthoryear{{Bocchino}, {Bandiera}, \&
  {Gelfand}}{{Bocchino} et~al.}{2010}]{2010A&A...520A..71B}
{Bocchino}, F., {Bandiera}, R.,  \& {Gelfand}, J. 2010, \aap, 520, A71

\bibitem[\protect\citeauthoryear{{de Jager} \& {Djannati-Ata{\"i}}}{{de Jager}
  \& {Djannati-Ata{\"i}}}{2008}]{2008arXiv0803.0116D}
{de Jager}, O.~C.,  \& {Djannati-Ata{\"i}}, A. 2008, ArXiv 0803.0116

\bibitem[\protect\citeauthoryear{{Dickey} \& {Lockman}}{{Dickey} \&
  {Lockman}}{1990}]{1990ARA&A..28..215D}
{Dickey}, J.~M.,  \& {Lockman}, F.~J. 1990, \araa, 28, 215

\bibitem[\protect\citeauthoryear{{Draine}}{{Draine}}{2003}]{2003ApJ...598.1026%
D}
{Draine}, B.~T. 2003, \apj, 598, 1026

\bibitem[\protect\citeauthoryear{{Gaensler} \& {Johnston}}{{Gaensler} \&
  {Johnston}}{1995}]{1995MNRAS.275L..73G}
{Gaensler}, B.~M.,  \& {Johnston}, S. 1995, \mnras, 275, L73

\bibitem[\protect\citeauthoryear{{Gallant} et~al.}{{Gallant}
  et~al.}{2008}]{2008AIPC..983..195G}
{Gallant}, Y.~A., et~al. 2008, in American Institute of Physics Conference
  Series, Vol. 983, 40 Years of Pulsars: Millisecond Pulsars, Magnetars and
  More, ed. {C.~Bassa, Z.~Wang, A.~Cumming, \& V.~M.~Kaspi}, 195

\bibitem[\protect\citeauthoryear{{Gelbord} et~al.}{{Gelbord}
  et~al.}{2010}]{2010GCN.10526....1G}
{Gelbord}, J.~M., et~al. 2010, GRB Coordinates Network, Circular Service,
  10526, 1, 526, 1

\bibitem[\protect\citeauthoryear{{Gelbord} \& {Vetere}}{{Gelbord} \&
  {Vetere}}{2010}]{2010GCN.10531....1G}
{Gelbord}, J.~M.,  \& {Vetere}, L. 2010, GRB Coordinates Network, Circular
  Service, 10531, 1, 531, 1

\bibitem[\protect\citeauthoryear{{G{\"o}{\u g}{\"u}{\c s}} et~al.}{{G{\"o}{\u
  g}{\"u}{\c s}} et~al.}{2010}]{2010ApJ...718..331G}
{G{\"o}{\u g}{\"u}{\c s}}, E., et~al. 2010, \apj, 718, 331

\bibitem[\protect\citeauthoryear{{Hinton} \& {Hofmann}}{{Hinton} \&
  {Hofmann}}{2009}]{2009ARA&A..47..523H}
{Hinton}, J.~A.,  \& {Hofmann}, W. 2009, \araa, 47, 523

\bibitem[\protect\citeauthoryear{{Kargaltsev} et~al.}{{Kargaltsev}
  et~al.}{2008}]{2008ApJ...684..542K}
{Kargaltsev}, O., {Misanovic}, Z., {Pavlov}, G.~G., {Wong}, J.~A.,  \&
  {Garmire}, G.~P. 2008, \apj, 684, 542

\bibitem[\protect\citeauthoryear{{Kargaltsev} \& {Pavlov}}{{Kargaltsev} \&
  {Pavlov}}{2008}]{2008AIPC..983..171K}
{Kargaltsev}, O.,  \& {Pavlov}, G.~G. 2008, in American Institute of Physics
  Conference Series, Vol. 983, 40 Years of Pulsars: Millisecond Pulsars,
  Magnetars and More, ed. {C.~Bassa, Z.~Wang, A.~Cumming, \& V.~M.~Kaspi}, 171

\bibitem[\protect\citeauthoryear{{Kargaltsev} \& {Pavlov}}{{Kargaltsev} \&
  {Pavlov}}{2010}]{2010AIPC.1248...25K}
{Kargaltsev}, O.,  \& {Pavlov}, G.~G. 2010, in American Institute of Physics
  Conference Series, Vol. 1248, American Institute of Physics Conference
  Series, ed. {A.~Comastri, L.~Angelini, \& M.~Cappi}, 25

\bibitem[\protect\citeauthoryear{{Kassim}}{{Kassim}}{1992}]{1992AJ....103..943%
K}
{Kassim}, N.~E. 1992, \aj, 103, 943

\bibitem[\protect\citeauthoryear{{LaMassa}, {Slane}, \& {de Jager}}{{LaMassa}
  et~al.}{2008}]{2008ApJ...689L.121L}
{LaMassa}, S.~M., {Slane}, P.~O.,  \& {de Jager}, O.~C. 2008, \apjl, 689, L121

\bibitem[\protect\citeauthoryear{{Landi} et~al.}{{Landi}
  et~al.}{2006}]{2006ApJ...651..190L}
{Landi}, R., et~al. 2006, \apj, 651, 190

\bibitem[\protect\citeauthoryear{{Mathis} \& {Lee}}{{Mathis} \&
  {Lee}}{1991}]{1991ApJ...376..490M}
{Mathis}, J.~S.,  \& {Lee}, C. 1991, \apj, 376, 490

\bibitem[\protect\citeauthoryear{{Mukherjee}, {Gotthelf}, \&
  {Halpern}}{{Mukherjee} et~al.}{2009}]{2009ApJ...691.1707M}
{Mukherjee}, R., {Gotthelf}, E.~V.,  \& {Halpern}, J.~P. 2009, \apj, 691, 1707 (M+09)

\bibitem[\protect\citeauthoryear{{O'C.~Drury} et~al.}{{O'C.~Drury}
  et~al.}{2009}]{2009A&A...496....1D}
{O'C.~Drury}, L., {Aharonian}, F.~A., {Malyshev}, D.,  \& {Gabici}, S. 2009,
  \aap, 496, 1

\bibitem[\protect\citeauthoryear{{Predehl} \& {Schmitt}}{{Predehl} \&
  {Schmitt}}{1995}]{1995A&A...293..889P}
{Predehl}, P.,  \& {Schmitt}, J.~H.~M.~M. 1995, \aap, 293, 889

\bibitem[\protect\citeauthoryear{{Smith} \& {Dwek}}{{Smith} \&
  {Dwek}}{1998}]{1998ApJ...503..831S}
{Smith}, R.~K.,  \& {Dwek}, E. 1998, \apj, 503, 831

\bibitem[\protect\citeauthoryear{{Tian} et~al.}{{Tian}
  et~al.}{2007}]{2007ApJ...657L..25T}
{Tian}, W.~W., {Li}, Z., {Leahy}, D.~A.,  \& {Wang}, Q.~D. 2007, \apjl, 657,
  L25

\bibitem[\protect\citeauthoryear{{Walter} et~al.}{{Walter}
  et~al.}{2006}]{2006A&A...453..133W}
{Walter}, R., et~al. 2006, \aap, 453, 133

\bibitem[\protect\citeauthoryear{{Weingartner} \& {Draine}}{{Weingartner} \&
  {Draine}}{2001}]{2001ApJ...548..296W}
{Weingartner}, J.~C.,  \& {Draine}, B.~T. 2001, \apj, 548, 296

\bibitem[\protect\citeauthoryear{{Yamazaki} et~al.}{{Yamazaki}
  et~al.}{2006}]{2006MNRAS.371.1975Y}
{Yamazaki}, R., {Kohri}, K., {Bamba}, A., {Yoshida}, T., {Tsuribe}, T.,  \&
  {Takahara}, F. 2006, \mnras, 371, 1975

\end{thebibliography}

\end{document}